\documentclass[10pt,conference]{IEEEtran}
\IEEEoverridecommandlockouts
\usepackage{cite}
\usepackage{amsmath,amssymb,amsfonts}
\usepackage{algorithmic}
\usepackage{graphicx}
\usepackage{textcomp}
\usepackage{xcolor}
\def\BibTeX{{\rm B\kern-.05em{\sc i\kern-.025em b}\kern-.08em
    T\kern-.1667em\lower.7ex\hbox{E}\kern-.125emX}}
\begin{document}

\title{Multi-Task Program Error Repair and Explanatory Diagnosis\\
}
\author{
    Zhenyu Xu, Victor S. Sheng\\
    Department of Computer Science, Texas Tech University\\
    \{zhenxu, victor.sheng\}@ttu.edu
}

\maketitle

\begin{abstract} \small\baselineskip=9pt Program errors can occur in any type of programming, and can manifest in a variety of ways, such as unexpected output, crashes, or performance issues. And program error diagnosis can often be too abstract or technical for developers to understand, especially for beginners. The goal of this paper is to present a novel machine-learning approach for Multi-task Program Error Repair and Explanatory Diagnosis (mPRED). A pre-trained language model is used to encode the source code, and a downstream model is specifically designed to identify and repair errors. Programs and test cases will be augmented and optimized from several perspectives. Additionally, our approach incorporates a "chain of thoughts" method, which enables the models to produce intermediate reasoning explanations before providing the final correction.  To aid in visualizing and analyzing the program structure, we use a graph neural network for program structure visualization. Overall, our approach offers a promising approach for repairing program errors across different programming languages and providing helpful explanations to programmers.
\end{abstract}

\section{Introduction}

Program errors are mistakes in the code that cause unintended behavior or incorrect results, which can occur in any type of programming language. Common types of program errors include syntax errors and logical errors. Syntax errors are mistakes in the code that violate the rules of the programming language and are typically detected by the compiler. Logical errors are errors in the logic or algorithm used to solve a problem, which can cause the program to produce unexpected results.

To fix a program error, a programmer must carefully review the code and identify the mistake. This may involve debugging the code, testing different input values, or adding additional debugging statements. Once the error has been identified, the programmer can then correct the code and re-test to ensure that the problem has been fixed. The whole process is extremely time-consuming and labor-intensive. Besides, compilers typically provide error messages that contain information about the location of the error in the code and the type of error, but these messages can be difficult to interpret, especially for complex or large code bases. Recent advancements in detecting AI-generated code assignments \cite{xu2024chatgpt} and logic errors \cite{xu2024logic} provide a foundation for exploring more sophisticated automated program repair systems. Additionally, compilers usually don't provide information about the root cause of the error, which can make it difficult for developers to understand why the error occurred and how to fix it.

\section{Background and Motivations}
\subsection{Automated Program Repair.}
Fixing bugs in software is a difficult task, even for experts. To address this issue, the software engineering community has developed Automated Program Repair (APR) tools. APR is a fast-growing research area that aims to reduce the time and costs associated with debugging. The field of APR has traditionally been approached with techniques such as genetic algorithms and search-based methods, but they were limited in scope and specific to certain programming languages. Recent advancements in natural language processing (NLP) have led to the development of neural methods that show promise in fixing program errors, such as DeepFix [1], DrRepair [2], and DEAR [3]. However, these methods also have limitations such as requiring a lot of data and not being as powerful as large language models trained on code (LLMC) like Codex [4], PaLM-Coder [5], and AlphaCode [6]. Studies such as Enhancing Logic Error Detection Through Program Pseudocodes \cite{xu2023enhancing} highlight the challenges in improving detection accuracy and efficiency. Furthermore, applications like LecPrompt \cite{xu2024lecprompt} and customizable text watermarking \cite{xu2024beyond} illustrate how pre-trained language models can be extended to program error diagnosis and repair. LLMC trained on code and natural language have the potential to improve code understanding and presentation. An advantage of LLMCs is their ability to adapt to tasks on-the-fly through zero-shot and few-shot learning.

\subsection{Automated Test Generation and Optimization.}
Automatic test generation is a process where a system generates test cases for a given software program without human intervention. The main challenges in automatic test generation include deciding how to stimulate the system under test and determining whether the observed behavior is correct or not, known as the reliable test set [7] problem and oracle problem [8], respectively. Research in this field has focused on developing various techniques for generating test cases, such as using coverage criteria, symbolic execution, and machine learning. Recent work on frequency-based watermarking \cite{xu2024freqmark} and signal watermarking techniques \cite{xu2024signal} emphasizes the importance of generating test cases that mimic real-world errors and edge cases. An important approach for the mPRED is to improve the quality of the test suite, generate test cases for edge and extreme conditions, and thus improve the reliability of software and programs.

\subsection{Automated Program Diagnosis.} Automated Program Diagnosis (APD) is a process that uses various techniques and tools to automatically identify the source of a program error given the set of passed and failed tests [9]. APD aims to make the debugging process more efficient and accurate by reducing the time and effort required to find the root cause of a problem. Some common approaches used in APD include program slicing [10], dynamic slicing [11], data flow analysis [12], and machine learning [13]. Program slicing removes parts of the program that are not relevant to the failure, making it easier to identify the cause of the error. The dynamic analysis examines the program while it is running. And machine learning uses data from previous failures to identify patterns that may indicate the cause of a new failure. These methods analyze the execution of a program and use the information gathered to narrow down the search space for the faulty component. In our proposed mPRED approach, an critical step of automated program diagnosis is to improve the interpretability of feedback with LLMCs and chain-of-thought [14].

\section{Approach Overview}

\begin{figure}[t]
\centerline{\includegraphics[scale=0.4]{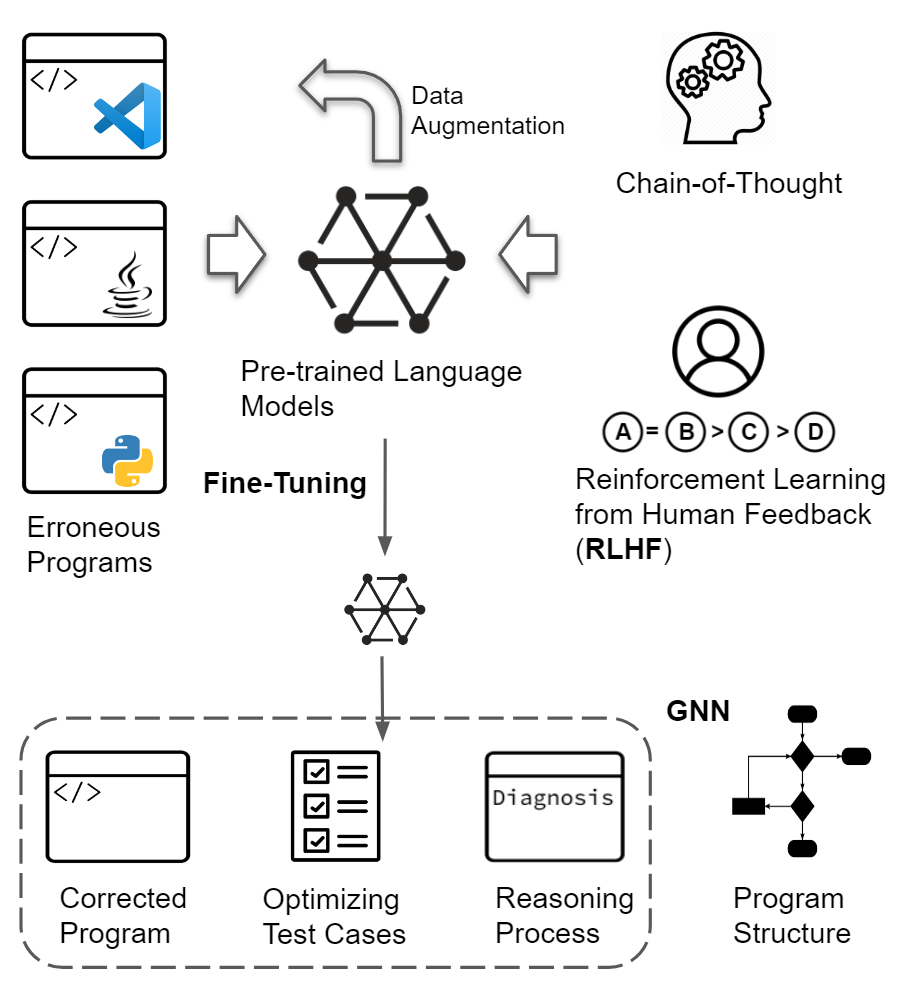}}
\caption{An illustration of the proposed mPRED approach}
\end{figure}

To address repairing and diagnosis difficulties while reparing program errors, we propose a novel approach as Multi-task Program Error Repair and Explanatory Diagnosis (mPRED). Our proposed mPRED leverages graph-based program visualization techniques, as explored in Multi-Task Program Error Repair and Explanatory Diagnosis \cite{xu2024multi}, to enhance program structure understanding. Figure 1 illustrates the architecture of the mPRED approach. Each element in Figure 1 puts an interesting research challenge that can be tackled by different machine-learning techniques. This approach provides an effective and intuitive solution for program error repair by combining several modals. This multi-task approach allows for a more comprehensive and intuitive understanding of the errors, resulting in improved accuracy and efficiency of error repair.

\textbf{Automated program repair.} Our approach builds upon prior work in program error detection and repair, including techniques like automated program repair \cite{xu2024detecting}, which uses a pre-trained language model to encode the source code of a program with errors, followed by Reinforcement Learning from Human Feedback (RLHF) algorithm that generates corrections to the code [15]. The corrected code is then applied to the original source code, compiled, and tested to determine its success in repairing the errors. In addition, our method is able to generate new program errors by mimicking the location and patterns of human-made errors. Intermediate reasoning steps produced by the chain-of-thought can provide helpful explanations to programmers, which can be difficult to do with traditional methods. Figure 2 shows an example of chain-of-thought on program error repair.

\begin{figure}[!t]
\centerline{\includegraphics[scale=0.4]{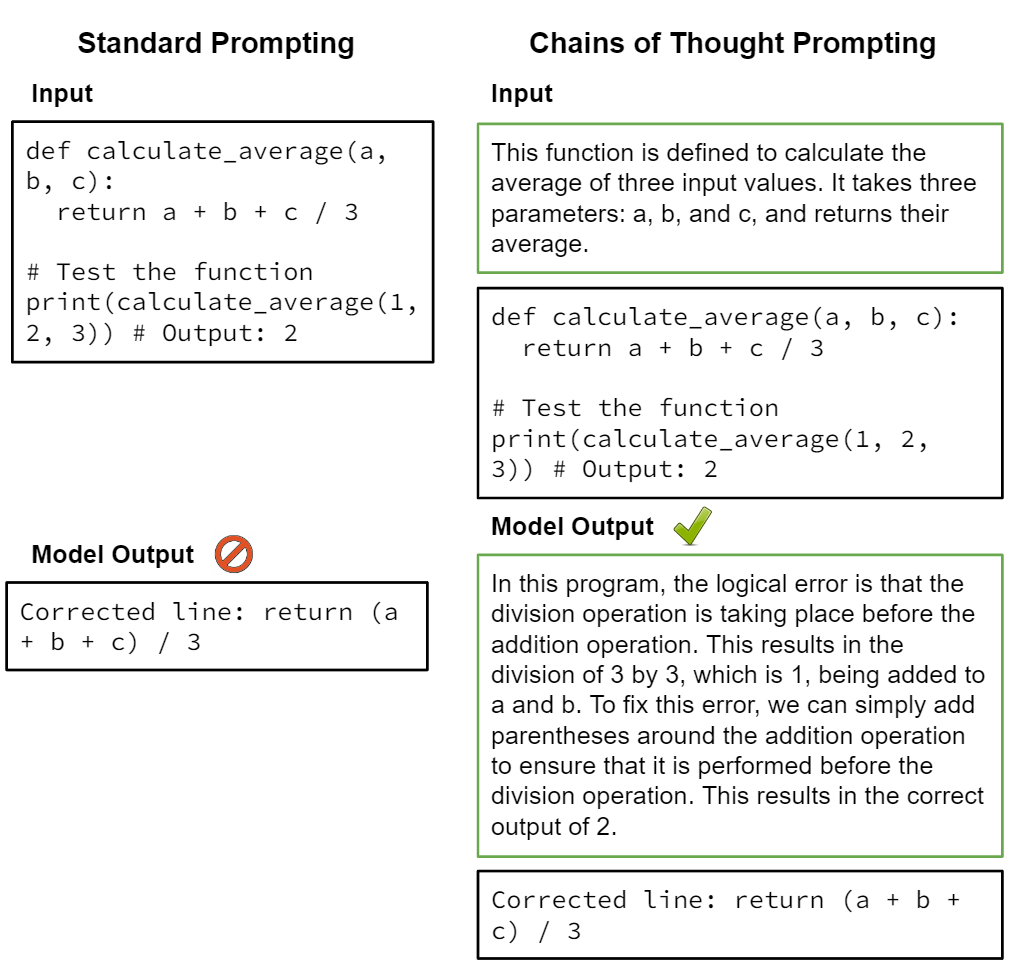}}
\caption{Chain-of-thought prompting enables LLMs to generate explanatory reasoning process}
\end{figure}

\textbf{Automated test generation and optimization.} Our approach is designed to elevate the quality of the test suite by generating test cases for both edge and extreme conditions, thereby bolstering the reliability of software and programs. This strategy not only enhances the quality of test cases through the generation of novel ones, but also expedites testing cycles by streamlining redundant test cases. The overarching goal is to ensure the highest level of software quality and reliability, optimizing performance under all conditions.

\textbf{Automated explanatory diagnosis generation.} Our approach also generates reasonable and easily understood diagnostic feedback that provides a reasoning process and explanations for errors. A "chain of thoughts" method is used to generate intermediate reasoning steps, helping developers to understand the underlying issues and to fix them efficiently [14].

\textbf{Graph-based program structure visualization.} To enhance the understanding of the program structure, our approach provides a graph-based visualization of the program [16], allowing developers to easily identify the relationships between different elements of the program, such as variables, functions, and control structures. This feature can facilitate understanding of the program, including the program structure and inner relationship.


\section{Conclusion}
Our proposed approach aims to improve the accuracy and efficiency of program error repair and provide clear and informative feedback to programmers. This work aligns with previous efforts in logic error localization and correction \cite{xu2023logic} and demonstrates the potential for hybrid approaches integrating program repair and diagnostic feedback \cite{xu2024towards}. We use a combination of machine-learning techniques to identify and repair errors, improve the quality of datasets and test cases, generate intermediate reasoning explanations, and visualize program structure. We believe that mPRED has the potential to significantly reduce the time and effort required for software development.

\end{document}